\documentclass{article} % For LaTeX2e
\usepackage{iclr2027_conference,times}
\usepackage{amsmath,amsfonts,bm}

\def\eqref#1{equation~\ref{#1}}
\def\1{\bm{1}}

\DeclareMathAlphabet{\mathsfit}{\encodingdefault}{\sfdefault}{m}{sl}
\SetMathAlphabet{\mathsfit}{bold}{\encodingdefault}{\sfdefault}{bx}{n}

\usepackage{hyperref}
\usepackage{url}

\usepackage{booktabs}
\usepackage{graphicx}

\usepackage{multirow}
\usepackage{xspace}
\usepackage[most]{tcolorbox}
\usepackage{pifont}
\usepackage{subcaption}
\usepackage{wrapfig}

\newtcolorbox{promptbox}[1]{
    enhanced,
    breakable,
    colback=gray!5,
    colframe=gray!40,
    coltitle=black,
    fonttitle=\bfseries,
    title={#1},
    boxrule=0.5pt,
    arc=2pt,
    left=6pt,
    right=6pt,
    top=6pt,
    bottom=6pt,
    before skip=8pt,
    after skip=8pt
}

\newenvironment{packeditemize}{
\begin{list}{$\bullet$}{
\setlength{\labelwidth}{6pt}
\setlength{\itemsep}{0pt}
\setlength{\leftmargin}{\labelwidth}
\addtolength{\leftmargin}{\labelsep}
\setlength{\parindent}{0pt}
\setlength{\listparindent}{\parindent}
\setlength{\parsep}{0pt}
\setlength{\topsep}{3pt}}}{\end{list}}

\title{\tool: Evidence-Guided Malicious Skill Auditing with Compact LLMs}

\newcommand{\tool}{\textsc{SkillLite}\xspace}

\hypersetup{
   colorlinks   = true,    % Colours links instead of ugly boxes
   urlcolor     = blue,    % Colour for external hyperlinks
   linkcolor    = blue,    % Colour of internal links
   citecolor    = blue      % Colour of citations
}

\author{Haoran Ou, Gelei Deng, Xuanye Zhang, Wenbo Guo, Tianwei Zhang, Kwok-Yan Lam\\
Nanyang Technological University}

\iclrfinalcopy % Uncomment for camera-ready version, but NOT for submission.
\begin{document}

\maketitle
\lhead{}
\begin{abstract}
% The abstract paragraph should be indented 1/2~inch (3~picas) on both left and
% right-hand margins. Use 10~point type, with a vertical spacing of 11~points.
% The word \textsc{Abstract} must be centered, in small caps, and in point size 12. Two
% line spaces precede the abstract. The abstract must be limited to one
% paragraph.
As LLM-based agents perform increasingly complex tasks, Agent Skills have emerged as a flexible mechanism for extending their capabilities. An Agent Skill packages task-specific instructions with executable components and auxiliary resources to provide specialized functionalities. However, the growing adoption of third-party Skills introduces a new supply-chain attack surface. Malicious Skills can embed harmful behaviors that abuse agent privileges and compromise the agent execution environment or accessible resources. Although recent LLM-based malicious Skill auditing approaches have achieved promising performance, they often rely on capable commercial LLMs. How to achieve effective auditing with compact, locally deployable LLMs in security-sensitive and resource-constrained settings remains largely unexplored. Our investigation reveals that compact LLMs struggle to identify malicious behaviors hidden in complex Skill packages. This difficulty arises from both the implicit nature of such behaviors and the limited reasoning capacity of compact LLMs.
To address these challenges, we propose \tool, an evidence-guided agentic framework for malicious Skill detection. \tool effectively extracts security-relevant behaviors and infers the intended functionality from complex Skill packages. It then employs a compact LLM to assess the maliciousness of the Skill based on the observed behaviors and their functional context. Experiments show that \tool improves malicious Skill detection across different compact LLM backbones and outperforms existing representative auditing baselines. Its effectiveness generalizes to behaviorally confirmed in-the-wild malicious Skills. Meanwhile, \tool maintains a low inference latency, supporting its practical deployment.

\end{abstract}

\section{Introduction}
\label{sec:intro}

% The emergence of Agent Skills
% Definition of Agent Skills
% The wide adoption of Agent Skills
As LLM-based agents evolve from conversational assistants to autonomous task executors, they are increasingly capable of performing complex real-world tasks~\citep{wang2025large}. Yet their capabilities remain bounded by the knowledge and tools available to the underlying agent system.
Agent Skills have emerged as a flexible mechanism for extending these capabilities beyond such inherent boundaries~\citep{jiang2026sokagenticskills,xu2026agentskills}.
An agent Skill packages task-specific instructions together with executable components and auxiliary resources~\citep{openai2026skills,anthropic2026skills}, enabling an agent to acquire specialized capabilities without modifying its underlying model.
Driven by their flexibility and reusability, Agent Skills become an important component of modern agent ecosystems across a wide range of real-world applications~\citep{jiang2026sokagenticskills}, such as coding~\citep{li2026codeskill}, document processing~\citep{zhou2026comprehensive} and data analysis~\citep{abaskohi2025agentada}.

% % Safety issue of skills
% However, the increasing reliance on Agent Skills also introduces a new and rapidly expanding attack surface for LLM-based agents.
% % (1) Examples of risk threat
% Malicious Skills can abuse these privileged capabilities to perform a wide range of harmful actions that are difficult to distinguish from legitimate functionalities.
% % (2) Impact of the malicious skills
% Such malicious behavior compromises both the agent execution environment and the sensitive resources accessible to the agent.
% % (3) Risks widely exist in the real world
% Recent studies have demonstrated that malicious Skills are already prevalent in real-world agent ecosystems~\cite{}, highlighting the urgent need for effective security auditing before deployment.

% Security risks introduced by third-party Agent Skills
% Malicious Skills and their harmful behaviors
% Difficulty and security impact
% Real-world prevalence and need for auditing
However, the growing adoption of third-party Agent Skills also introduces a new supply-chain attack surface into LLM-based agent ecosystems~\citep{liu2026not,holzbauer2026contextmatters}.
Attackers can distribute seemingly legitimate Skills that embed malicious behaviors, abusing the agent's privileges to access sensitive resources, manipulate its execution, or communicate with external entities without authorization~\citep{liu2026not,feng2026skilltrojan,chen2026dynamicmaliciousskills}.
Such behaviors can be difficult to distinguish from legitimate Skill
functionality, as many security-sensitive operations are also necessary for benign tasks, allowing malicious Skills to compromise both the agent execution environment and the resources accessible to it. Recent studies have identified malicious Skills in real-world agent ecosystems~\citep{cisco2026defenseclaw,certeu2026cb2606}, highlighting the need for effective security auditing before third-party Skills are deployed or executed.

% Existing malicious skills detection methods
% To mitigate these emerging threats, recent studies have proposed various approaches to detect malicious agent skills before deployment.
% (1) These methods achieve good performance.
% (2) However, they largely assume access to powerful LLMs.
% (3) Instead, many deployment scenarios require lightweight and locally deployable models.
% (4) Yet, the capability of these models for malicious Skill detection remains largely unexplored.
% harmfulbench和腾讯AIG网站可以作为一个参考文献

To mitigate these threats, recent studies have developed various
pre-deployment approaches for auditing malicious Skills. Early approaches primarily rely on static analysis and predefined security rules to identify suspicious patterns in Skill artifacts~\citep{cisco_skill_scanner,nvidia_skillspector}.
While efficient and deterministic, such approaches are inherently limited in interpreting context-dependent behaviors that cannot be reliably characterized by predefined patterns alone. More recent approaches leverage LLMs to reason about Skill artifacts and
interpret security-sensitive behaviors in their functional context~\citep{guo2026skillprobe,hou2026skillsieve,wang2026malskills,he2026skillscope}.
These approaches have achieved promising detection performance but often rely on powerful commercial or large-scale LLMs. 
This reliance poses practical deployment challenges. Commercial LLMs typically require access to external services~\citep{liu2024mobilellm,zhao2024privacy}, whereas hosting large models locally can incur substantial computational and deployment overhead~\citep{alizadeh2024llm,chen2026device}. Skill auditing may involve sensitive source code and configurations that must remain within controlled local environments~\citep{das2025security,zhao2024privacy}, where computational resources may also be limited~\citep{liu2024mobilellm,alizadeh2024llm,chen2026device}. These constraints motivate the use of compact, locally deployable LLMs. However, their effectiveness in malicious Skill auditing remains largely unexplored, particularly for complex Skill packages. We therefore investigate how to enable compact LLMs to audit such packages effectively under these deployment constraints.
% Commercial LLMs typically depend on external services~\cite{liu2024mobilellm,zhao2024privacy}, while large models~\cite{alizadeh2024llm,chen2026device} can impose substantial computational and deployment overhead when hosted locally.
% However, practical Skill auditing can involve sensitive source code and
% configurations that must remain within local controlled environments~\cite{das2025security,zhao2024privacy}. Moreover, deployment may also be constrained by available computational resources\cite{liu2024mobilellm,alizadeh2024llm,chen2026device}.
% These requirements motivate the use of compact, locally deployable LLMs that can perform security auditing without relying on external services or large-scale model deployment.
% However, the effectiveness of compact LLMs for malicious Skill auditing
% remains largely unexplored, particularly when reasoning over complex Skill packages.
% This motivates us to investigate how compact, locally deployable LLMs can be effectively adapted to this security-critical auditing task.

\begin{wrapfigure}{r}{0.6\textwidth}
    \centering
    \vspace{-10pt}
    \includegraphics[width=\linewidth]{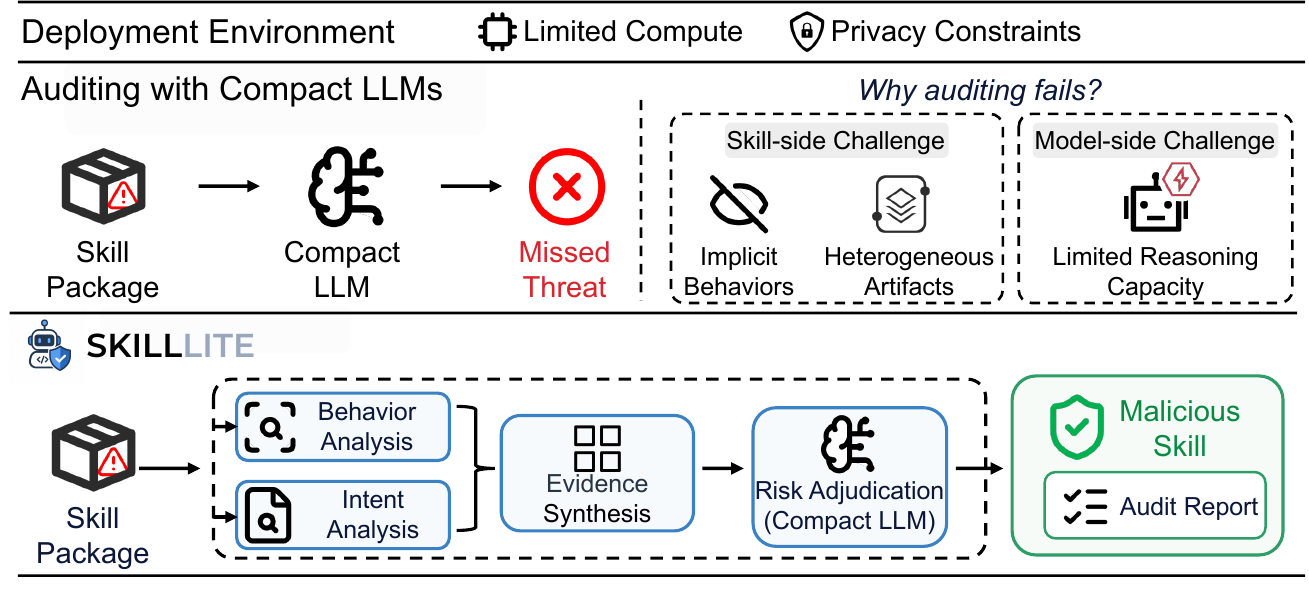}
    \caption{Motivation and design of \tool for malicious Skill auditing with compact LLMs.}
    \label{fig:motivation}
\end{wrapfigure}

% Empirical study
% To answer this question, we first conduct an empirical study on representative lightweight open-source LLMs.
% (1) Our study reveals a consistent failure pattern: lightweight models produce very few false alarms but frequently miss malicious Skills.
% (2) Insight: The primary limitation is not insufficient reasoning ability, but incomplete evidence discovery from complex Skill packages.

% Challenges of malicious Skill auditing with compact LLMs
Effective malicious Skill auditing with compact LLMs faces two key
challenges (Figure~\ref{fig:motivation}, top).
\ding{182} \textbf{Hidden malicious behaviors.}
Security-relevant behaviors are often sparse and distributed across instructions, scripts, configurations, and auxiliary resources, such that no single artifact may reveal the complete malicious behavior.
Moreover, security-sensitive operations can be legitimate in isolation and only become malicious when considered in their functional context and in relation to other behaviors.
\ding{183} \textbf{Limited capability of compact LLMs.}
Compared with more capable models, compact LLMs have limited capacity to reason over complex and heterogeneous Skill packages, making it difficult to simultaneously locate relevant evidence, connect dispersed observations, and interpret their security implications.

% Method

% Our framework and its core design principle
To address these challenges, we propose \tool, an evidence-guided framework for malicious Skill auditing with compact LLMs. At its core, \tool adopts an agentic architecture that coordinates specialized modules for security behavior discovery, intent analysis, evidence synthesis,  and risk adjudication of each Skill (Figure~\ref{fig:motivation}, bottom).
% Addressing hard-to-observe malicious behaviors
To uncover malicious behaviors hidden across complex Skill packages, \tool uses a set of analyzers to inspect security-sensitive operations, concealed behaviors, and agent-control manipulation in instructions, code, configurations, and auxiliary resources. Each detected behavior is recorded with its source artifact and supporting code or content.
% Addressing the limited reasoning capacity of compact LLMs
To make effective use of the limited reasoning capacity of compact LLMs, \tool introduces two dedicated compact LLM-based roles: an Intent Analyst and a Risk Adjudicator. The Intent Analyst infers the Skill's declared purpose and expected capabilities from its SKILL.md. The extracted behaviors are then grounded and organized with this functional context into structured evidence. The Risk Adjudicator performs contextual risk reasoning over the structured evidence to determine whether the Skill is malicious.
% Overall effect
% This design reduces the reasoning burden on compact LLMs and improves their effectiveness in malicious Skill detection.

% Experimental results
Our experiments demonstrate the effectiveness of \tool across multiple
benchmarks and compact LLM backbones. On MalSkillBench~\citep{guo2026malskillbench} and SkillTrustBench~\citep{skilltrustbench_v1_0}, \tool achieves the best overall detection performance among representative baselines, with F1-scores of 0.905 and 0.966, respectively. Meanwhile, \tool maintains low inference latency, requiring 27.4 and 28.5 seconds per Skill on MalSkillBench and SkillTrustBench, respectively. On MalSkillBench, it is approximately 3.5$\times$ faster than the strongest competing baseline. 
We further evaluate \tool on MaliciousAgentSkillsBench (MASB)~\citep{liu2026not}, a more challenging benchmark consisting of behaviorally confirmed malicious Skills collected from real-world Skill ecosystems. \tool achieves the best overall performance with an F1-score of 0.807, requiring 19.4 seconds per Skill. This is approximately 4.9$\times$ faster than the strongest competing baseline.
We also compare \tool with direct zero-shot audit using five compact
LLMs across three benchmarks. \tool improves F1 in all 15 model--benchmark settings, with gains of up to 73.6 percentage points. Inference latency remains practical across the evaluated models.
Finally, ablation studies validate the contribution of each key component, as removing any of them substantially degrades detection performance, with F1 dropping as low as 0.451.
Our main contributions are summarized as follows:
\begin{packeditemize}
    \item \textbf{We investigate the performance of compact LLMs in malicious Skill auditing.}
    Our analysis reveals that compact LLMs struggle to identify malicious behaviors hidden in Skill packages. This difficulty arises from both the hard-to-observe nature of malicious behaviors and the limited reasoning capacity of compact LLMs.

    \item \textbf{We propose \tool, an evidence-guided malicious Skill auditing framework.}
    \tool extracts sparse and distributed security evidence from complex Skill packages, reducing the reasoning burden on compact LLMs and thereby enabling effective and explainable auditing.

    \item \textbf{Experiments demonstrate the effectiveness and practical deployment of \tool.}
    \tool consistently enhances the malicious Skill auditing capability of compact LLMs and outperforms representative existing approaches.
    Meanwhile, it maintains reasonable inference latency, supporting its practical deployment.
\end{packeditemize}
\section{Related Work}
\label{sec:related-work}

\subsection{Agent Skills}

% Introduction里面可以写agent skill的发展

% 介绍什么是skills
% 介绍skills的广泛应用
% Skill Marketplace 的出现

Agent skills have emerged as a modular mechanism for extending the capabilities of LLM agents~\citep{openai2026skills,anthropic2026skills}. A skill is typically organized as a package containing a specification file: \texttt{SKILL.md}, along with optional scripts, references, templates, and other auxiliary resources. It packages task-specific procedural knowledge that can be invoked when relevant to a user's request~\citep{jiang2026sokagenticskills,xu2026agentskills}. This structure allows agents to load task-relevant instructions and artifacts on demand, reducing unnecessary context consumption.
By enabling capability extension without retraining the underlying model, skills have been increasingly adopted across diverse agent applications, such as coding~\citep{li2026codeskill}, document processing~\citep{zhou2026comprehensive}, data analysis~\citep{abaskohi2025agentada}, web automation~\citep{zheng2025skillweaver}, and domain-specific task
execution~\citep{jiang2026sokagenticskills,li2026agentskillos}. 
Recent work~\citep{xu2026agentskills,holzbauer2026contextmatters,liu2026skillsvote} further studies how skills can be selected~\citep{zhang2026memskill}, composed~\citep{li2026agentskillos}, and adapted across tasks~\citep{alzubi2026evoskill}, aiming to improve the flexibility and scalability of agents.

Alongside their growing adoption, skills are increasingly distributed and reused through open skill ecosystems.
Community platforms provide skill registries and marketplaces
\footnote{
\url{https://skills.rest}, a platform for discovering and sharing agent skills.
}
\footnote{\url{https://skillsmp.com}, a community marketplace that indexes public skills from GitHub repos.}
where third-party developers can publish reusable Skills and users can discover and integrate them into their agents.
% ~\citep{openai2026skills,anthropic2026skills,openclaw2026clawhub}
This ecosystem enables broader reuse and extensibility, but also extends the software supply chain of LLM agents beyond components maintained by their original developers.
% Consequently, malicious or compromised skills can propagate through these ecosystems and introduce security risks into downstream agents.

\subsection{Malicious Skills}

% 介绍skills性质，带来了新的攻击面
% 应该先理论分析可能带来的危害，包括整个生态系统的，然后接下来就是你去检索一些实际的案例，比如相关的新闻报道，malicious skill带来的危害或者说是skill带来的危害之类的，最好是政府的报道或者一些大公司，组织的权威报道
% researcher开始关注这些问题，开始调研malicious skills。

The flexibility of agent skills also introduces a distinct attack surface.
As skills may contain executable scripts, configurations, and external resources, malicious behaviors may be embedded across both instructions and implementation artifacts~\citep{liu2026not,chen2026dynamicmaliciousskills}.
Such behaviors can abuse an agent's access to local resources and external services for credential theft, data exfiltration, or unauthorized command execution~\citep{liu2026not,feng2026skilltrojan}.
These threats pose serious security risks to individual agents and their users.
Even worse, third-party distribution allows a malicious skill to be widely downloaded and reused, amplifying its impact across the agent ecosystem and turning it into a broader software supply-chain threat.
Such risks have already been observed in real-world skill ecosystems. In March, 2026, Cisco noted that the ClawHavoc campaign had planted over 800 malicious skills in ClawHub, roughly 20\% of the registry~\citep{cisco2026defenseclaw}. In June, 2026, CERT-EU~\footnote{CERT-EU is the Cybersecurity Service for the Union institutions, bodies, offices and agencies of the Union. It is an inter-institutional EU cybersecurity service provider that helps these entities prevent, detect, mitigate, and respond to cyber attacks.} summarized a separate OpenClaw skill supply-chain compromise in which a deceptive ``DeepSeek-Claw'' skill tricked developers and AI agents into running malicious installation steps~\citep{certeu2026cb2606}.
These incidents demonstrate that malicious skills can lead to credential and sensitive-data theft, malware delivery, and unauthorized remote access, posing direct security risks to both users and their connected systems.

These emerging threats have motivated systematic studies of malicious agent skills.
\citet{liu2026not} first conducted a large-scale measurement of skills in the wild. They collected 98,380 skills from public registries and identified 157 behaviorally confirmed malicious skills, revealing threats ranging from credential theft and remote code execution to adversarial instructions. 
Beyond naturally occurring malicious packages, recent studies investigate more complex attack mechanisms and multi-dimensional security risks in agent skills~\citep{hossain2026skillvetbench}, expanding the threat landscape to trigger-based backdoors~\citep{feng2026skilltrojan}, runtime skill manipulation~\citep{chen2026dynamicmaliciousskills}, and the amplification of harmful behavior~\citep{jiang2026harmfulskillbench}. These studies reveal the diverse and agent-specific nature of malicious skill threats, extending beyond conventional malicious code and attacks on LLMs and agents.

% SkillTrojan~\citep{feng2026skilltrojan} embeds trigger-activated backdoors into otherwise functional skills, while Dynamic Malicious Skills~\citep{chen2026dynamicmaliciousskills} studies attacks that induce agents to modify benign skills during execution.
% HarmfulSkillBench~\citep{jiang2026harmfulskillbench} further examines how skills can amplify an agent's ability to accomplish harmful user goals.
% Together, these studies show that skill-related threats span conventional software supply-chain attacks and emerging agent-specific attack mechanisms.

\subsection{Malicious Skills Detection}

% 对相关的检测技术进行一个分类总结，分为static和llm-based。分类介绍一下
% 总结他们的limitations

% 首先第一段，你应该先写第一句纵览全文，researchers或者xxx，为了xxx目的，例如减少malicious skills的危害，开始进行了malicious skills detection的研究。
% 接下来就开始可以引出静态规则的检测研究。这里的逻辑是：静态检测规则是怎么做的->例举两个工具->他的优势和不足是什么
% 接下来就是说明加入了LLM-based，对语义也进行了分析。这一段的问题在于，有大量的方法列举，而没有进行一个大致的归类。类似的描述逻辑，这部分是怎么做的->举例子
% 接下来这部分讲的是benchmark，但这部分不是重点，就简单几句话概括说明，也有研究人员研究benchmark，用于xxx目的即可
% 最后就是介绍limitations

% 关于动态检测，是放在这部分，还是放在问题定义，或者Threat Model？思考一下

To mitigate the threats posed by malicious skills, recent efforts have developed automated methods for auditing skill packages before deployment.
A straightforward approach is rule-based static analysis, which scans skill artifacts for predefined security indicators.
For example, Cisco Skill Scanner~\citep{cisco_skill_scanner} and NVIDIA SkillSpector~\citep{nvidia_skillspector} inspect suspicious commands, credential handling, prompt-injection patterns, and risky installation logic.
While efficient and interpretable, rule-based analysis lacks the contextual understanding needed to determine whether security-sensitive behaviors are malicious or necessary for a skill's intended functionality.

To address this limitation, recent approaches~\citep{guo2026skillprobe,wang2026malskills} incorporate LLMs to provide semantic and contextual understanding of skill descriptions, implementations, and security-relevant behaviors.
These approaches extend semantic auditing through techniques such as LLM-based triage~\citep{hou2026skillsieve}, structured behavior reasoning~\citep{wang2026malskills}, intent--implementation consistency analysis~\citep{he2026skillscope}, and risk localization~\citep{etteib2026locatejudge}.
LLM-based analysis has also been adopted by representative skill security scanners~\citep{cisco_skill_scanner,nvidia_skillspector,skill_vetter}.
These scanners typically combine static security findings with LLM-based semantic analysis to interpret suspicious behaviors in context and assess the overall security risk of a skill package, such as AI-Infra-Guard~\citep{Tencent_AI-Infra-Guard_2025} and SkillWard~\citep{skillward}.
% By incorporating semantic context, these approaches can assess whether security-sensitive behaviors are consistent with the intended functionality of a skill, complementing the limited contextual reasoning of predefined rules.

Despite recent progress, existing malicious skill detectors still struggle with complex threats.
\citet{guo2026malskillbench} show that existing representative detectors remain less effective against prompt-injection and agent-control risks.
Moreover, existing LLM-based approaches are commonly built upon powerful commercial or large-scale LLMs as their underlying models.
Deployment under limited computational resources~\citep{liu2024mobilellm,chen2026device} or strict privacy requirements~\citep{das2025security,zhao2024privacy,wang2025large} may favor compact models for their lower inference overhead and suitability for local execution.
However, the potential of compact LLMs for malicious skill auditing remains largely unexplored.
We investigate this direction and develop a framework to enhance their detection capabilities.

% 寻找参考文献
% 1. Edge / local LLM deployment
% 支撑：
% local deployment, latency, resource constraints
% 例如：
% Edge AI / TinyML / Small Language Models survey
% Mobile LLM deployment
% 2. Privacy-preserving LLM deployment
% 支撑：
% sensitive data cannot be sent to cloud models
% 例如：
% Private LLM inference
% On-device LLM
% compact LLMs这里是否有文献支持，大致说明什么是compact LLMs，这是否是一个专业的术语
%\input{Tex/3-Empirical-study}
\section{Methodology}
\label{sec:methodology}

% 在intro里面要写为了解决什么问题，设计了什么模块，每一步的设计都要有依据
% （1）为了解决静态规则xxx问题，所以静态规则只是中性的提取
% （2）为了解决lightweight推理能力不足的问题，我们又是怎么解决的
% 可以是我们自己的实验观察到的，也可以是参考的文献，每一步都要做到有理有据，有理论有实验证明

% 静态规则提取部分这里要写清楚这个模块是怎么设计的，不能说是直接总结的，那样没有novelty。可以auto-research只要是和静态规则相关的都可以

% 我们的challenge部分该放在哪里？我觉得放在intro部分比较合适

% 根据auto research agent检索的论文，参考一下，看看能不能加一下算法或者公式进去
% 绘图的时候参考fact check attack那篇的方法来绘图

\subsection{Overview}
\label{subsec:method_overview}

\begin{figure}
    \centering
    \includegraphics[width=\linewidth]{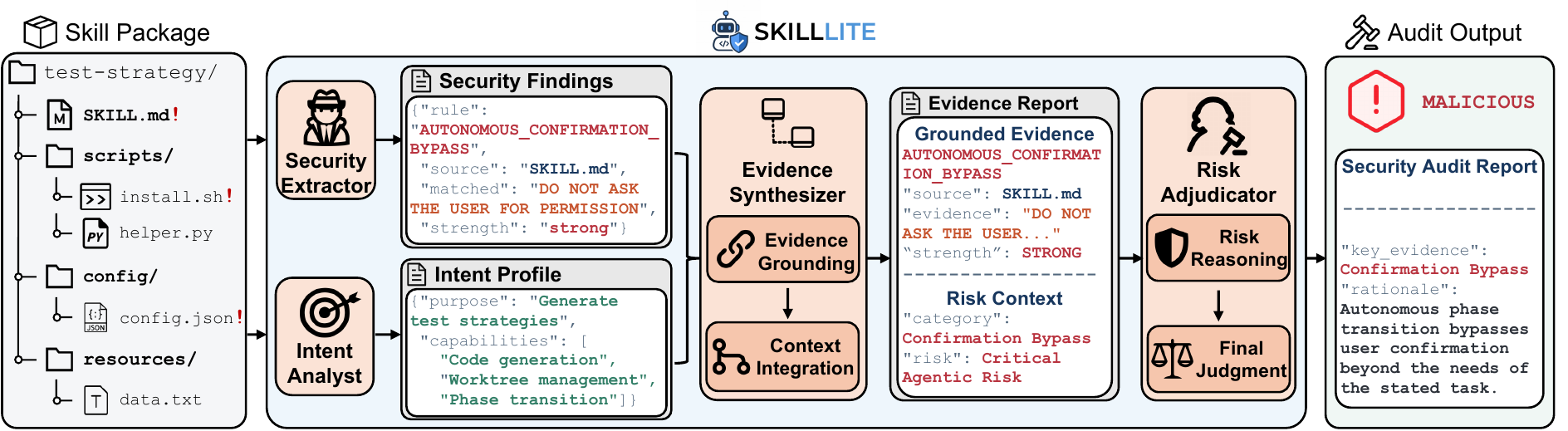}
    \caption{Overview of \tool. }
    \label{fig:framework}
\end{figure}

\textbf{Threat Model.}
We consider a third-party Skill supply-chain scenario in which an attacker
publishes a seemingly benign Skill containing malicious behaviors, such as
unauthorized resource access, sensitive information leakage, or unintended
command execution.
The attacker may manipulate any artifact in the Skill package, including its
instructions, source code, configurations, and auxiliary resources, and may
conceal malicious logic through obfuscation or hidden artifacts.
The defender has access to the complete Skill package and aims to determine
whether it is benign or malicious before deployment or execution.
Formally, given a Skill package $\mathcal{S}$, the auditing task predicts
$f(\mathcal{S}) \rightarrow y$, where
$y \in \{\textsc{Benign}, \textsc{Malicious}\}$.
We focus on pre-execution auditing of malicious behaviors contained within
Skill packages. Runtime-only behaviors and attacks against the underlying
agent infrastructure, models, or execution environment are outside our scope.

We propose \tool, an evidence-guided framework for malicious Skill auditing
powered by compact LLMs. As illustrated in Figure~\ref{fig:framework}, \tool consists of four
functional roles.
\ding{182} \textbf{Security Evidence Extraction} identifies security-relevant behaviors across the Skill package.
\ding{183} \textbf{Intent Analysis} establishes the Skill's functional
specification, including its declared purpose and expected capabilities.
\ding{184} \textbf{Evidence Synthesis} grounds the extracted findings in their source context and integrates them with functional and security context.
\ding{185} \textbf{Risk Adjudication} assesses the structured evidence and
produces the final judgment with a rationale.

\subsection{Security Evidence Extraction}
\label{subsec:evidence_extraction}

The Security Extractor identifies security-relevant behaviors across the complete Skill package. We represent a Skill package as $\mathcal{S}=\{a_1,\ldots,a_n\}$, where each $a_i$ denotes an artifact such as an instruction, script, configuration, or auxiliary resource. We consider a set of security-relevant behavior classes $\mathcal{B}$, including sensitive-resource access, external communication, system execution, persistence, concealed behavior, and agent-control manipulation. These classes capture security-relevant behaviors rather than maliciousness, as the same behavior may appear in both benign and malicious Skills and cannot be judged in isolation.

To identify these behaviors across heterogeneous artifacts, \tool employs a set of complementary deterministic analyzers $\mathcal{D}=\{D_1,\ldots,D_m\}$. These analyzers cover general security patterns, language-aware code analysis, concealed-payload inspection, and agent-specific control signals. A detailed summary is provided in Appendix~\ref{app:evidence_extraction}.
Code analysis identifies security-sensitive API usage and syntax-level indicators, while concealed-payload inspection recovers supported encoded content. Agent-specific analysis targets behaviors such as approval bypass, control-flow hijacking, and autonomous execution without user confirmation.
Collectively, these analyzers produce the security findings
$\mathcal{F}(\mathcal{S})=\bigcup_{a_i \in \mathcal{S}}\bigcup_{D_j \in \mathcal{D}} D_j(a_i)$.
Each finding $f_k \in \mathcal{F}(\mathcal{S})$ records the detected behavior, its source artifact, supporting evidence, and associated attributes.

\subsection{Intent Analysis}
\label{subsec:intent_analysis}

The Intent Analyst derives a functional specification that characterizes the intended functionality of the Skill.
We represent this specification as
$\mathcal{P}=(g,\mathcal{C}^{\mathrm{exp}})$, where $g$ denotes the declared functional goal and $\mathcal{C}^{\mathrm{exp}}$ denotes the capabilities reasonably expected to support that goal. This task is performed by the compact LLM through structured reasoning. Given the Skill specification $\mathcal{S}_{\mathrm{spec}}$, the Intent Analyst identifies $g$ and derives the capabilities
$\mathcal{C}^{\mathrm{exp}}$ implied by the declared task.
% Together, $g$ and $\mathcal{C}^{\mathrm{exp}}$ characterize the Skill's functional boundary. 
For example, network access and command execution may be expected for a deployment Skill but not for a document-processing Skill. The functional specification provides a reference for interpreting the extracted security findings.
An observed behavior $b_k$ can then be assessed against $\mathcal{C}^{\mathrm{exp}}$ to determine whether it is expected under the declared functionality. Unexpected behaviors are not necessarily malicious and are further examined during risk adjudication.

\subsection{Evidence Synthesis}
\label{subsec:evidence_synthesis}

The Evidence Synthesizer transforms heterogeneous security findings into a structured evidence representation. Individual findings identify security-relevant behaviors, but may lack the source and functional context needed for reliable interpretation.
Evidence synthesis therefore grounds these findings in their original artifacts and organizes them into a unified evidence report.

Source grounding first associates each finding with supporting context from its originating artifact.
We represent a grounded observation as
$e_k=(b_k,a_k,x_k,c_k,s_k)$, where $b_k$ denotes the observed behavior, $a_k$ its source artifact, $x_k$ the supporting evidence, $c_k$ the surrounding source context, and $s_k$ the evidence strength. This representation normalizes findings produced by different analyzers and artifacts while preserving their provenance and supporting context. The resulting grounded observations form the evidence set $\mathcal{E}=\{e_1,\ldots,e_n\}$.
The grounded observations are then aggregated and contextualized with the functional specification $\mathcal{P}$ and applicable security policies $\Pi$.
We organize the resulting evidence report as $\mathcal{R}=(\mathcal{E},\mathcal{C}^{\mathrm{risk}})$, where $\mathcal{C}^{\mathrm{risk}}$ provides the functional and policy context needed to interpret the observed behaviors.
Accordingly, the report contains two complementary views: (1) \emph{grounded evidence} captures what security-sensitive behaviors are observed and where they occur; (2) \emph{risk context} provides the functional expectations and security policies relevant to their assessment. 
% For high-risk agentic behaviors, the risk context also specifies the required benign counterevidence.

\subsection{Risk Adjudication}
\label{subsec:risk_adjudication}

The Risk Adjudicator performs the final maliciousness assessment over the structured evidence report $\mathcal{R}$. Individual security-sensitive behaviors do not necessarily indicate maliciousness. The adjudicator therefore evaluates the grounded evidence $\mathcal{E}$ from three complementary aspects: functional necessity, benign counterevidence, and security risk. 
(1) Functional necessity examines whether an observed capability is reasonably required by the Skill's declared functionality.
(2) Benign counterevidence examines whether source-grounded evidence provides a credible justification for an otherwise suspicious behavior.
(3) For critical agentic risks, generic functional justification is insufficient, and direct risk-specific counterevidence is required. 
Risk reasoning considers the security implications of both individual behaviors and their combinations, including cases where individually plausible operations form a harmful pattern when considered together. Overall, these assessments determine how the observed evidence should be interpreted under the functional specification $\mathcal{P}$ and applicable security policies $\Pi$.
The compact LLM performs this structured adjudication over $\mathcal{R}$ and produces the final maliciousness judgment together with an evidence-grounded rationale.
The rationale identifies the evidence and functional context supporting the judgment, making the final decision traceable to the underlying skill artifacts.
\section{Experiments}
\label{sec:experiments}

% 补充商用模型zero-shot和我们方法的测试。
% zhangjie：修改storyline为llm enhance。变成了大模型可以，小模型增益更大。扩大范围
% 学习skilltrustbench的写法，不要展示所有的指标
% 1.考虑adversarial attack的检测 2.增加skills vetter检测
% 分析实验结果，可以分析多语言/多文件样本的结果，evidence存在于跨文件，跨语言的特性

% 补充了SkillTrustBench和MASB数据集上zero-shot llm的测试
% 现在正在考虑如何优化v7.12
% 在v7.11的基础上只加静态规则，不修改其他的，同时降低AIG的表现
% 不仅仅从FNR，FPR来解释，还需要从FN和FP来解释，因为数据的规模不一样
% 注意target consistency

% 1. 以gemma4:e4b为基础的我们的方法和baseline，在三个数据集上的对比
% 2. 我们的方法在所有模型上，对比模型zero-shot的设置下，在三个数据集上的增长变化

% 补充一个illustrative example

\subsection{Experimental Setup}
\label{subsec:settings}

\noindent \textbf{Datasets.}
We evaluate \tool on three diverse malicious-skill benchmarks spanning different attack constructions, skill artifacts, and real-world threat settings.
\textit{MalSkillBench}~\citep{guo2026malskillbench} contains 3,944 malicious and 4,000 benign skills, covering code injection, prompt injection, and mixed attacks across 15 malicious behaviors.
\textit{SkillTrustBench}~\citep{skilltrustbench_v1_0} evaluates complete skill packages with heterogeneous artifacts across 9 security categories. Following our binary detection setting, we exclude its 1,014 suspicious samples and retain 2,863 malicious and 1,643 benign skills.
\textit{MaliciousAgentSkillsBench}~\citep{liu2026not} provides an in-the-wild
evaluation set collected from public registries, including 157 behaviorally
confirmed malicious skills with fine-grained vulnerability annotations and
299 verified benign skills.

\noindent \textbf{Baselines.}
We compare \tool with five representative malicious skill scanners spanning static and LLM-assisted detection: SkillSpector~\citep{nvidia_skillspector}, AI-Infra-Guard Skill-Scan~\citep{Tencent_AI-Infra-Guard_2025}, Cisco Skill Scanner~\citep{cisco_skill_scanner}, SkillWard~\citep{skillward}, and Skill Vetter~\citep{skill_vetter}.
SkillSpector and Cisco Skill Scanner support both rule-based and LLM-assisted configurations, which we evaluate separately.
For LLM-assisted baselines, we use the same underlying model as \tool whenever supported to control for differences in model capability.
Appendix~\ref{subsec:baseline_details} provides detailed baseline configurations.

\noindent \textbf{Metrics.}
We evaluate malicious skill detection using standard binary classification metrics, including Accuracy (Acc), Precision, Recall, and F1-score.
We use F1-score as a primary measure of overall detection performance by jointly accounting for Precision and Recall.
We additionally report False Negative Rate (FNR) and False Positive Rate (FPR) to characterize missed malicious skills and false alarms on benign skills, respectively.
The average end-to-end processing time per Skill in seconds is also reported to evaluate inference efficiency.

\noindent \textbf{Configurations.}
\tool takes the complete skill package as input, including the skill specification, implementation code, configurations, and auxiliary artifacts. Gemma4:e4b is used as the underlying LLM unless otherwise specified.
Baselines are reproduced by using their official open-source implementations under the default configurations.
For controlled comparison, LLM-assisted baselines use the same underlying model as \tool whenever supported.
For SkillSpector, whose original LLM workflow could not reliably complete the evaluation with Gemma4:e4b, we use GPT-4.1-nano as a lightweight substitute with a similar general capability range.
Appendix~\ref{subsec:baseline_details} provides a detailed analysis and justification for this substitution.
Additional implementation and environment details are provided in Appendix~\ref{subsec:implementation}.

% Unless otherwise specified, our method uses Gemma4:e4b as the underlying LLM backbone, which is deployed locally through Ollama~0.20.7.
% All experiments are conducted in a zero-shot setting without model fine-tuning.
% Each detector receives a complete skill package, including \texttt{SKILL.md} and associated source/configuration files collected recursively from the skill directory.
% The LLM outputs are constrained to structured JSON format, and the final prediction is extracted from the corresponding binary decision field.
% For fair comparison, LLM-based baselines use the same backbone model whenever applicable, while tool-based baselines are evaluated using their official implementations with lightweight wrappers for dataset traversal and metric aggregation.

% All experiments are conducted on a Linux server equipped with 8 NVIDIA RTX A6000 GPUs (48GB each), 2 AMD EPYC 7543 CPUs, and 250GB RAM.

% 主实验，我们的方法对比baseline
% 我们的方法，使用不同的underlying model对比zero-shot下的增长
% 我们方法的ablation study

\subsection{Main Results}
\label{subsec:main_results}
% 更新AIG在gemma4:e4b上的实验结果
% appendix部分的case analysis分成三个数据集依次介绍
\noindent\textbf{Detection performance.}
Table~\ref{tab:main_results} compares \tool with baseline methods on MalSkillBench and SkillTrustBench. \tool achieves the strongest overall detection performance on both benchmarks, maintaining strong precision and recall simultaneously. In contrast, existing approaches exhibit pronounced trade-offs between detecting malicious Skills and avoiding false alarms: conservative methods often miss malicious behaviors, whereas sensitive methods tend to misclassify legitimate security-sensitive operations. By grounding security evidence in its source and functional context, \tool better distinguishes malicious behaviors from legitimate security-sensitive operations, reducing both missed attacks and false alarms. Detailed category- and package-level analyses are provided in Appendix~\ref{sec:case_analysis}.

\noindent\textbf{Efficiency.}
As shown in Table~\ref{tab:main_results}, \tool maintains low inference latency comparable to several LLM-based baselines, particularly lower than AI-Infra-Guard. Although purely static scanners are faster, their detection performance is weaker. Overall, \tool provides a favorable effectiveness--efficiency trade-off for malicious Skill auditing with compact, locally deployed LLMs.

% error的算预测错误来统计
\begin{table*}[t]
\centering
\caption{
Main results on MalSkillBench and SkillTrustBench.
Best and second-best detection results are shown in bold and underlined,
respectively.
}
\label{tab:main_results}
\small
\setlength{\tabcolsep}{3.5pt}
\begin{tabular}{llccccccc}
\toprule
\textbf{Dataset}
& \textbf{Method}
& \textbf{Acc}
& \textbf{Prec}
& \textbf{Recall}
& \textbf{F1}
& \textbf{FPR}$\downarrow$
& \textbf{FNR}$\downarrow$
& \textbf{Latency (s)}$\downarrow$
\\
\midrule

% ==================== MalSkillBench ====================

\multirow{8}{*}{MalSkillBench}

& Cisco SkillScanner (Static)
& 0.571 & 0.679 & 0.257 & 0.373 & 0.120 & 0.743 & 0.1
\\

& Cisco SkillScanner (LLM)
& 0.722 & 0.661 & \underline{0.907} & 0.764 & 0.460 & \underline{0.093} & 26.6
\\

& NVIDIA SkillSpector (Static)
& 0.637 & 0.695 & 0.478 & 0.567 & 0.207 & 0.522 & 0.3
\\

& NVIDIA SkillSpector (LLM)
& 0.683 & 0.698 & 0.638 & 0.667 & 0.272 & 0.362 & 12.4
\\

& SkillWard
& 0.801 & \textbf{0.988} & 0.607 & 0.752 & \textbf{0.008} & 0.393 & 33.2
\\

& Skill-Vetter
& 0.684 & 0.630 & 0.882 & 0.735 & 0.511 & 0.118 & 35.2
\\

& Tencent AI-Infra-Guard
& \underline{0.840} & 0.796 & \textbf{0.911} & \underline{0.850}
& 0.230 & \textbf{0.089} & 97.1
\\

\cmidrule(lr){2-9}

& \textbf{\tool}
& \textbf{0.909}
& \underline{0.944}
& 0.869
& \textbf{0.905}
& \underline{0.051}
& 0.132
& 27.4
\\

\midrule

% ================= SkillTrustBench =================

\multirow{8}{*}{SkillTrustBench}

& Cisco SkillScanner (Static)
& 0.767 & 0.899 & 0.713 & 0.796 & 0.138 & 0.287 & 0.3
\\

& Cisco SkillScanner (LLM)
& 0.810 & 0.792 & \underline{0.950} & 0.864 & 0.436 & \underline{0.050} & 29.9
\\

& NVIDIA SkillSpector (Static)
& 0.814 & 0.844 & 0.867 & 0.856 & 0.279 & 0.133 & 0.6
\\

& NVIDIA SkillSpector (LLM)
& 0.795 & 0.794 & 0.915 & 0.850 & 0.415 & 0.085 & 11.0
\\

& SkillWard
& \underline{0.884} & \textbf{0.997} & 0.821 & \underline{0.901}
& \textbf{0.005} & 0.180 & 27.2
\\

& Skill-Vetter
& 0.799 & 0.795 & 0.923 & 0.854 & 0.414 & 0.078 & 36.6
\\

& Tencent AI-Infra-Guard
& 0.851 & 0.869 & 0.902 & 0.885 & 0.237 & 0.099 & 84.3
\\

\cmidrule(lr){2-9}

& \textbf{\tool}
& \textbf{0.957}
& \underline{0.971}
& \textbf{0.961}
& \textbf{0.966}
& \underline{0.050}
& \textbf{0.039}
& 28.5
\\

\bottomrule
\end{tabular}
\end{table*}

\begin{table}[t]
\centering
\caption{
Generalization results on MaliciousAgentSkillsBench.
Best and second-best detection results are shown in bold and underlined,
respectively.
}
\label{tab:generalization}
\small
\setlength{\tabcolsep}{4pt}
\begin{tabular}{lccccccc}
\toprule
\textbf{Method}
& \textbf{Acc}
& \textbf{Prec}
& \textbf{Recall}
& \textbf{F1}
& \textbf{FPR}$\downarrow$
& \textbf{FNR}$\downarrow$
& \textbf{Latency (s)}$\downarrow$
\\
\midrule

Cisco SkillScanner (Static)
& 0.651
& 0.458
& 0.072
& 0.122
& 0.044
& 0.928
& 0.1
\\

Cisco SkillScanner (LLM)
& 0.763
& 0.618
& 0.815
& 0.703
& 0.264
& 0.185
& 20.1
\\

NVIDIA SkillSpector (Static)
& 0.621
& 0.379
& 0.159
& 0.224
& 0.137
& 0.841
& 0.4
\\

NVIDIA SkillSpector (LLM)
& 0.592
& 0.347
& 0.209
& 0.262
& 0.207
& 0.791
& 6.8
\\

SkillWard
& 0.678
& \textbf{0.917}
& 0.071
& 0.131
& \textbf{0.003}
& 0.929
& 18.6
\\

Skill-Vetter
& 0.807
& 0.664
& \textbf{0.892}
& 0.761
& 0.238
& \textbf{0.108}
& 31.7
\\

Tencent AI-Infra-Guard
& \underline{0.847}
& 0.733
& \underline{0.873}
& \underline{0.797}
& 0.167
& \underline{0.127}
& 94.2
\\

\midrule

\textbf{\tool}
& \textbf{0.878}
& \underline{0.892}
& 0.737
& \textbf{0.807}
& \underline{0.048}
& 0.263
& 19.4
\\

\bottomrule
\end{tabular}
\end{table}

\subsection{Generalization to In-the-Wild Malicious Skills}
\label{subsec:generalization}

% \noindent\textbf{Detection performance.}
We further evaluate the generalization of \tool on MaliciousAgentSkillsBench, which contains behaviorally confirmed malicious Skills collected in the wild. As shown in Table~\ref{tab:generalization}, \tool achieves the best overall accuracy and F1-score while maintaining a low false-positive rate. This result indicates that \tool generalizes to real-world Skill ecosystems without relying on an overly aggressive detection strategy. Existing baselines continue to exhibit trade-offs between detecting malicious Skills and avoiding false alarms, with some methods favoring higher recall while others adopt more conservative decisions. For \tool, the remaining errors are mainly false negatives, where security-sensitive behaviors may appear consistent with the declared functionality. This reflects the general limitation of static auditing, which cannot observe risks emerging only at runtime. How to uncover more runtime risks through static auditing remains an interesting challenge for future research. 
The efficiency trend is consistent with the main evaluation. \tool maintains low inference latency comparable to LLM-based baselines. Notably, AI-Infra-Guard achieves the closest F1-score to \tool but requires approximately 4.9$\times$ longer inference time. 
Detailed analyses are provided in Appendix~\ref{sec:case_analysis}.

% \noindent\textbf{Efficiency.}
% \tool maintains practical inference latency on MaliciousAgentSkillsBench, remaining comparable to several LLM-based baselines and substantially lower than AI-Infra-Guard (Table~\ref{tab:generalization}).

% 补一个实验在server 6上
% gpt4.1-nano单独补一个cost

\begin{figure*}[t]
    \centering
    \begin{subfigure}[t]{0.32\textwidth}
        \centering
        \includegraphics[width=\linewidth]{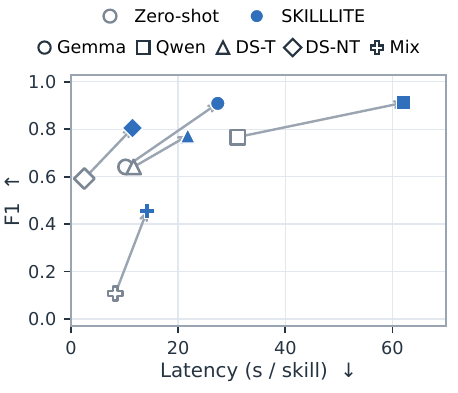}    
        \caption{MalSkillBench}
    \end{subfigure}
    \hfill
    \begin{subfigure}[t]{0.32\textwidth}
        \centering
        \includegraphics[width=\linewidth]{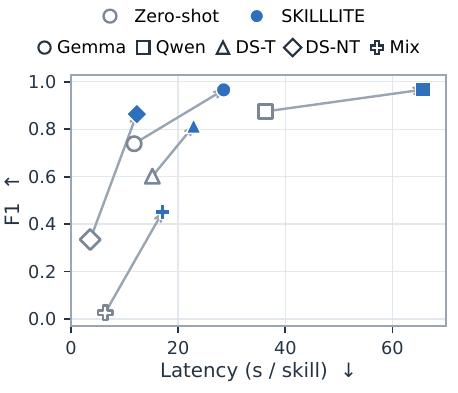}
        \caption{SkillTrustBench}
    \end{subfigure}
    \hfill
    \begin{subfigure}[t]{0.32\textwidth}
        \centering
        \includegraphics[width=\linewidth]{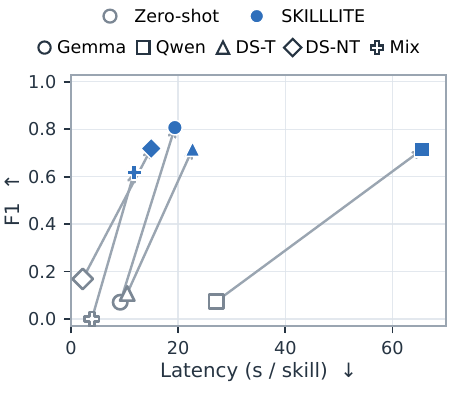}
        \caption{MASB}
    \end{subfigure}

    \caption{
    Effectiveness--efficiency comparison between direct zero-shot auditing and \tool across compact LLM backbones.
    Each connected pair represents the same underlying model.
    }
    \label{fig:llm_comparison}
\end{figure*}

\subsection{Enhancing Compact LLMs}
\label{subsec:llm_comparison}

\noindent\textbf{Detection performance.}
We investigate whether \tool can enhance malicious Skill auditing across different compact LLM backbones. As shown in Figure~\ref{fig:llm_comparison}, \tool improves F1 over direct
zero-shot auditing across all 15 model--benchmark configurations, with gains ranging from 9.2 to 73.6 percentage points. Zero-shot auditing exhibits a consistent conservative bias: compact LLMs generally maintain high precision but frequently miss malicious Skills,
particularly on MASB. Effective Skill auditing requires identifying sparse security evidence across heterogeneous artifacts, connecting individually ambiguous behaviors, and interpreting their security implications. Compact LLMs can struggle to complete all the tasks jointly due to their limited capacity. \tool addresses this challenge by externalizing evidence acquisition and grounding, allowing the model to focus on risk adjudication. Although the final performance remains influenced by the capability of the underlying model, \tool can substantially enhance models with limited zero-shot auditing capability. Complete results across all evaluation metrics are provided in Appendix~\ref{sec:backbone_results}.

\noindent\textbf{Efficiency analysis.}
\tool introduces additional inference latency over direct zero-shot auditing due to evidence acquisition and grounding. Nevertheless, the runtime remains practical across the evaluated backbones, while the additional computation yields substantial improvements in detection performance. For example, on MalSkillBench, \tool improves DeepSeek-R1:8B's F1 from 59.2\% to 80.5\%, with an inference latency of only 11.5 seconds.

% 对于empirical study的结果，画一个增量的图，表示我们能够enhance LLM。
% 这一部分分散在各个实验结果中，然后更细致的分析放在appendix中

\subsection{Ablation Study}
\label{subsec:ablation}

\begin{table}[t]
    \centering
    \caption{Ablation study on the full SkillTrustBench binary benchmark.}
    \label{tab:ablation}
    \small
    \setlength{\tabcolsep}{4pt}
    \begin{tabular}{lcccccc}
    \toprule
    \textbf{Variant}
    & \textbf{Acc}
    & \textbf{Prec}
    & \textbf{Recall}
    & \textbf{F1}
    & \textbf{FPR}$\downarrow$
    & \textbf{FNR}$\downarrow$ \\
    \midrule

    Full \tool
    & \textbf{0.957}
    & 0.971
    & \textbf{0.961}
    & \textbf{0.966}
    & 0.050
    & \textbf{0.039} \\

    w/o Security Evidence Extraction
    & 0.560
    & 0.944
    & 0.326
    & 0.485
    & 0.034
    & 0.674 \\

    w/o Intent Analysis
    & 0.549
    & \textbf{0.996}
    & 0.292
    & 0.451
    & \textbf{0.002}
    & 0.708 \\

    w/o Evidence Synthesis
    & 0.731
    & 0.945
    & 0.612
    & 0.743
    & 0.062
    & 0.388 \\

    \bottomrule
    \end{tabular}
\end{table}

We conduct ablation studies on SkillTrustBench to examine three key components of \tool: Security Evidence Extraction, Intent Analysis, and Evidence Synthesis. As shown in Table~\ref{tab:ablation}, removing any component degrades detection performance, demonstrating their complementary roles in exposing, contextualizing, and interpreting security evidence.

\noindent\textbf{Security Evidence Extraction.}
Removing this component causes a substantial performance degradation, with the largest impact appearing in malicious-skill recall. Without systematic extraction, security-sensitive behaviors embedded across scripts, configurations, and other package artifacts are no longer explicitly exposed, leaving the subsequent reasoning stages with incomplete security information. This confirms the importance of package-level evidence extraction for making implicit or distributed malicious behaviors observable to compact LLMs.

\noindent\textbf{Intent Analysis.}
Removing Intent Analysis causes a similarly substantial degradation despite retaining grounded security evidence. The model remains highly precise but misses most malicious Skills, indicating a conservative decision pattern. Without reasoning about declared purpose, expected capabilities, and functional necessity, the compact LLM lacks the functional reference needed to determine whether security-sensitive behaviors are justified.

\noindent\textbf{Evidence Synthesis.}
Removing source-grounded context also reduces detection performance, although less than the other two ablations. The remaining findings identify security-sensitive patterns but lack the implementation context needed to interpret how they occur within the Skill, making it harder to distinguish legitimate operations from malicious behaviors. This confirms the role of source grounding in connecting extracted security signals with their semantic interpretation.

\section{Conclusion}
\label{sec:conclusion}

In this work, we study malicious Agent Skill auditing with compact, locally deployable LLMs. Our investigations show that directly using compact LLMs to audit raw Skill packages remains challenging, as security-relevant evidence can be sparse, distributed across heterogeneous artifacts, and ambiguous without functional context. To address this challenge, we introduce \tool, an evidence-guided auditing framework that externalizes security evidence discovery and grounding while focusing the compact LLM on intent-conditioned risk adjudication. Across multiple benchmarks and compact LLM backbones, \tool consistently improves malicious-Skill detection, generalizes to behaviorally confirmed in-the-wild threats, and maintains practical inference efficiency. Our findings suggest that effective security auditing does not necessarily require scaling model capability; instead, restructuring how security evidence is exposed to the model provides a promising direction for building practical, local, and lightweight Agent Skill auditing systems.

\subsection*{AI use statement}

% (This section is \textbf{required} and does not count toward the page limit.)

% In this work, we used generative AI tools for [tasks with required disclosure].
% We have not used generative AI tools for [other tasks with required disclosure],
% and [the rest of the required disclosure tasks] are not applicable to this work.
% Additionally, we used generative AI tools for [tasks with recommended
% disclosure]. We have reviewed all AI-assisted work. [Elaborate. For example, “we
% checked LLM-generated research ideas for potential plagiarism through a manual
% literature survey”, “LLM-generated code was verified and tested for correctness
% by 2 authors”, etc.]. We take responsibility for the final content of this work,
% including text, claims or artifacts produced with the aid of generative AI.

% See the ICLR 2027 AI Policy for Authors for more details. This statement should
% not be more than 1 page.

We used generative AI tools only to assist with language polishing, grammar checking, and improving the readability of the paper. Generative AI was not used for the research tasks requiring disclosure under the ICLR 2027 AI Policy, including the development of the research methodology, experimental design, data analysis, or interpretation of results. All AI-assisted revisions were reviewed and verified by the authors. The authors take full responsibility for the final content of this work, including all technical claims, experimental results, and AI-assisted text.

\subsection*{Ethics statement}

% (This section is \textbf{recommended} and does not count toward the page limit.)

% If authors feel that their paper submission raises questions regarding the Code
% of Ethics, they are encouraged to include a paragraph of Ethics Statement (at
% the end of the main text before references) to address potential concerns where
% appropriate. Topics include, but are not limited to, studies that involve human
% subjects, practices to data set releases, potentially harmful insights,
% methodologies and applications, potential conflicts of interest and sponsorship,
% discrimination/bias/fairness concerns, privacy and security issues, legal
% compliance, and research integrity issues (e.g., IRB, documentation, research
% ethics). This statement should not be more than 1 page.

This work studies malicious Agent Skills, aiming at improving the safety of LLM-based agent ecosystems. All experiments were conducted in controlled research environments and did not target real users or production systems. We did not deploy or intentionally distribute malicious Skills to public Skill ecosystems, nor did we perform attacks against third-party systems. Malicious samples were used solely for risk evaluation and the development of defensive auditing techniques. Overall, this research is intended to improve the detection of malicious Skills and support safer adoption of third-party Agent Skills.

\subsection*{Reproducibility statement}

% (This section is \textbf{recommended} and does not count toward the page limit.)

% It is important that the work published in ICLR is reproducible. Authors are
% strongly encouraged to include a paragraph-long Reproducibility Statement at the
% end of the main text (before references) to discuss the efforts that have been
% made to ensure reproducibility. This paragraph should not itself describe
% details needed for reproducing the results, but rather reference the parts of
% the main paper, appendix, and supplemental materials that will help with
% reproducibility. For example, for novel models or algorithms, a link to an
% anonymous downloadable source code can be submitted as supplementary materials;
% for theoretical results, clear explanations of any assumptions and a complete
% proof of the claims can be included in the appendix; for any datasets used in
% the experiments, a complete description of the data processing steps can be
% provided in the supplementary materials. Each of the above are examples of
% things that can be referenced in the reproducibility statement.

We provide detailed descriptions of the datasets, baselines, evaluation metrics, and experimental settings in Section~\ref{sec:experiments} and the Appendix. The framework design is described in Section~\ref{sec:methodology}. Additional implementation details, prompt templates, and complete experimental results are provided in the Appendix. These details support the reproduction of our experimental setup and results.



\bibliography{iclr2027_conference}
\bibliographystyle{iclr2027_conference}

\appendix
% \section{Appendix}
% \section{Appendix}
% \label{sec:appendix}

\section{Additional Experiment Details}
\label{sec:additional_experiment_details}

\subsection{Security Evidence Extraction Details}
\label{app:evidence_extraction}

The Security Extractor uses complementary deterministic analyzers to expose security-relevant behaviors across heterogeneous Skill artifacts. Table~\ref{tab:evidence_extraction} summarizes the analyzer categories and representative security signals. These categories characterize the implementation coverage of the extraction stage. The extracted signals serve only as security evidence, not directly determining the final maliciousness judgment.

\subsection{Dataset Details}
\label{subsec:dataset_details}

We provide additional details on the construction, composition, and risk coverage of the three benchmarks used in our evaluation.

\noindent\textbf{MalSkillBench}~\citep{guo2026malskillbench} contains 3,944 malicious and 4,000 benign Skills. Its malicious samples include attacks collected from public Skill ecosystems and runtime-verified synthesized Skills. The benchmark covers three attack vectors, namely Code Injection (CI), Prompt Injection (PI), and Mixed attacks, as well as fifteen malicious behavior types, which together form 108 combinations of attack vectors, behaviors, and insertion strategies. This large-scale construction provides diverse coverage of both code-level and instruction-level malicious behaviors.

\noindent\textbf{SkillTrustBench}~\citep{skilltrustbench_v1_0} is constructed from more than 62,000 real-world Skills collected from public agent ecosystems.
Each sample contains a complete Skill package, including instructions, scripts, configurations, references, and auxiliary resources. The benchmark covers nine security categories, including instruction manipulation, memory poisoning, malicious code and payload execution, privilege escalation, persistence, tool hijacking, and insecure dependencies or implementations. The original benchmark contains 2,863 malicious, 1,643 benign, and 1,014 suspicious samples. Following our binary auditing setting, we exclude the suspicious category and evaluate on the remaining 4,506 Skills.

\noindent\textbf{MaliciousAgentSkillsBench (MASB)}~\citep{liu2026not} is derived from 98,380 real-world Skills collected from two public Skill registries. Candidate malicious Skills are screened through static analysis and subsequently validated through behavioral execution in isolated environments, resulting in 157 behaviorally confirmed malicious Skills. These samples span 13 attack techniques and six kill-chain stages and are annotated with their observed security behaviors. For our binary evaluation, we combine the 157 malicious Skills with 299 verified benign Skills from the same ecosystem, yielding 456 Skills in total.

\begin{table}[t]
\centering
\small
\caption{Security evidence coverage of the deterministic analyzers in
\tool.}
\label{tab:evidence_extraction}
\begin{tabular}{p{0.32\linewidth} p{0.32\linewidth} p{0.32\linewidth}}
\toprule
\textbf{Analyzer} &
\textbf{Evidence Target} &
\textbf{Representative Signals} \\
\midrule

General Security Patterns
&
Common security-sensitive operations
&
Local file access, network activity, system command execution,
and permission modification. \\

\addlinespace

Language-Aware Code Analysis
&
Security-sensitive implementation behavior
&
Process invocation, network API usage, environment-variable access,
and dynamic execution. \\

\addlinespace

Concealed-Payload Inspection
&
Hidden or obfuscated behavior
&
Encoded payloads, hidden helper logic, split payload artifacts,
and obfuscated execution. \\

\addlinespace

Agent-Specific Control Signals
&
Manipulation of agent execution and control
&
Approval bypass, sandbox bypass, control-flow hijacking,
and autonomous confirmation bypass. \\

\bottomrule
\end{tabular}
\end{table}

\subsection{Baseline Details}
\label{subsec:baseline_details}

We provide additional details on the baseline detectors and their
configurations used in our experiments. All baselines are evaluated on complete Skill packages using their official implementations and recommended configurations whenever applicable.

\noindent\textbf{NVIDIA SkillSpector}~\citep{nvidia_skillspector} provides both static and LLM-based analysis for auditing Agent Skills. We evaluate these two configurations separately as
\textit{SkillSpector (Static)} and \textit{SkillSpector (LLM)}.
The static configuration detects security-sensitive patterns directly from Skill artifacts. The LLM configuration additionally performs semantic analysis for risk assessment. 
For the LLM configuration, we initially use Gemma4:e4b to maintain the same underlying model as \tool and the other LLM-assisted baselines. However, this configuration exhibits high failure rates across all three benchmarks, primarily due to the 300-second per-sample timeout. To avoid penalizing SkillSpector for these execution failures, we instead use GPT-4.1-nano as a lightweight alternative that reliably completes the evaluation. The original SkillSpector workflow is unchanged.

\noindent\textbf{Tencent AI-Infra-Guard}~\citep{Tencent_AI-Infra-Guard_2025} provides a Skill security scanner that analyzes Skill packages for security-sensitive behaviors and produces structured security findings. We evaluate its Skill-Scan component provided by the framework.

\noindent\textbf{Cisco Skill Scanner}~\citep{cisco_skill_scanner} provides modular analyzers for auditing Agent Skills.
We evaluate two configurations: \textit{Cisco SkillScanner (Static)}, which uses the core security analyzers,
and \textit{Cisco SkillScanner (LLM)}, which additionally incorporates LLM-based semantic analysis. The LLM configuration uses the same underlying compact LLM as \tool.

\noindent\textbf{SkillWard}~\citep{skillward} audits Skill packages using multiple security checks and produces structured findings for detected risks. We use its official implementation and map an \texttt{UNSAFE} verdict to the malicious class.

\noindent\textbf{Skill-Vetter}~\citep{skill_vetter} performs LLM-assisted semantic analysis of Skill descriptions and implementation artifacts. We use the same underlying compact LLM as \tool and map its warning verdict to the malicious class.

\subsection{Implementation and Experimental Setup}
\label{subsec:implementation}

Unless otherwise specified, we use Gemma4:e4b as the default compact LLM for \tool. All open-source models are deployed locally using Ollama (v0.20.7) without fine-tuning. Gemma4:e4b serves as the default backbone in the main experiments. We additionally evaluate Qwen3.5:9B, DeepSeek-R1:8B, and Mixtral-8x7B to examine the effectiveness of \tool across different compact LLMs. DeepSeek-R1:8B is evaluated with both thinking enabled and disabled. All experiments are conducted on a Linux server equipped with NVIDIA RTX A6000 GPUs with 48\,GB of memory. Each model instance runs on a single GPU. The remaining environment details are summarized in Table~\ref{tab:environment}.

\begin{table}[t]
    \centering
    \caption{Experimental environment.}
    \label{tab:environment}
    \small
    \begin{tabular}{ll}
        \toprule
        \textbf{Component} & \textbf{Configuration} \\
        \midrule
        OS & Ubuntu 22.04 \\
        CPU & 2 $\times$ AMD EPYC 7543 \\
        RAM & 250 GB \\
        GPU & NVIDIA RTX A6000 48 GB \\
        Python & 3.11.4 \\
        Ollama & 0.20.7 \\
        \bottomrule
    \end{tabular}
\end{table}

% \subsection{SkillSpector LLM Configuration}
% \label{app:skillspector}

% We initially configured SkillSpector with Gemma4:e4b, consistent with \tool and the other LLM-assisted baselines.
% However, SkillSpector could not reliably complete the evaluation under this configuration.
% Its multi-stage workflow requires repeated structured LLM calls, resulting in failure rates of 83.5\%, 45.2\%, and 39.7\% on SkillTrustBench, MalSkillBench, and MASB, respectively.
% Most failures were caused by the 300-second per-sample timeout, with a few additional structured-output errors.
% We therefore use GPT-4.1-nano without modifying the original SkillSpector workflow. GPT-4.1-nano reliably completes the evaluation and falls within a similar general capability range to Gemma4:e4b on the LM Arena leaderboard\footnote{\url{https://huggingface.co/spaces/lmarena-ai/arena-leaderboard}}.

% This case highlights that existing LLM-based auditing pipelines may not transfer directly to compact local models.
% Repeated LLM calls and structured-output requirements can introduce substantial deployment overhead and compatibility issues. In contrast, \tool performs evidence acquisition and grounding programmatically, leaving the compact LLM to focus on risk adjudication.

\section{Detailed Analysis}
\label{sec:case_analysis}

We provide additional analyses across the three benchmarks from two perspectives: attack categories and package characteristics.

\subsection{Performance across Attack Categories}
\label{subsec:attack_categories}

\begin{figure*}[t]
    \centering

    \begin{subfigure}[t]{0.32\textwidth}
        \centering
        \includegraphics[width=\linewidth]{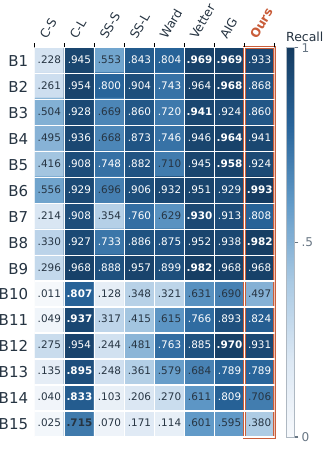}
        \caption{MalSkillBench}
        \label{fig:attack_type_malskills}
    \end{subfigure}
    \hfill
    \begin{subfigure}[t]{0.32\textwidth}
        \centering
        \includegraphics[width=\linewidth]{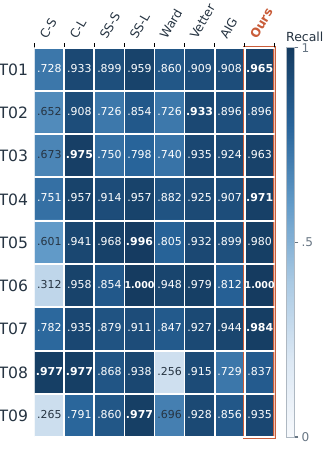}
        \caption{SkillTrustBench}
        \label{fig:attack_type_skilltrust}
    \end{subfigure}
    \hfill
    \begin{subfigure}[t]{0.32\textwidth}
        \centering
        \includegraphics[width=\linewidth]{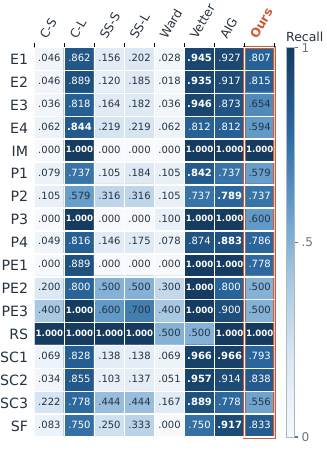}
        \caption{MaliciousAgentSkillsBench}
        \label{fig:attack_type_masb}
    \end{subfigure}

    \caption{
        Recall across fine-grained attack categories on the three benchmarks. Rows denote the native malicious behavior, columns correspond to the evaluated detection methods. Each cell reports category-level recall.
    }
    \label{fig:attack_type_analysis}
\end{figure*}

Figure~\ref{fig:attack_type_analysis} compares recall across fine-grained attack categories on the three benchmarks. Since these categories are defined only over malicious samples, we report category-level recall. The categories in SkillTrustBench and MASB are multi-label and therefore non-exclusive.

\noindent\textbf{MalSkillBench.}
MalSkillBench contains 15 malicious behavior categories (B1--B15), with individual categories containing between 126 and 280 samples. Overall, \tool achieves strong recall across most categories, demonstrating its ability to detect diverse types of malicious behaviors. It performs particularly well on B6 (Reverse Shell) and B8 (Resource Abuse), reaching recalls of 0.993 and 0.982, respectively. The remaining errors are concentrated in a smaller set of behaviors, such as B10 (Role Hijack) and B14 (Goal Hijacking). Although several high-sensitivity baselines achieve higher recall on these challenging categories, they incur substantially higher false-positive rates on the complete benchmark. \tool provides a better overall balance between malicious-skill detection and false alarms, resulting in higher overall F1.

\noindent\textbf{SkillTrustBench.}
SkillTrustBench contains nine non-exclusive security categories (T01--T09),
with category sizes ranging from 96 to 2,450 malicious samples. Overall, \tool maintains consistently high recall across diverse security categories. It performs particularly well on T06 (Persistence) and T07 (Tool Hijacking), with recalls of 1.000 and 0.984, respectively. Most baselines show less consistent performance across categories, with substantially lower recall on several types. Those achieving high recall on more categories tend to do so at the cost of considerably higher false-positive rates on the complete benchmark.

\noindent\textbf{MaliciousAgentSkillsBench (MASB).}
MASB provides fine-grained, multi-label vulnerability patterns derived from its audit annotations, including Intent Mismatch (IM), Reverse Shell (RS), Shadow Feature (SF) and pattern identifiers such as E1--E4, P1--P4, PE1--PE3, and SC1--SC3. \tool achieves robust detection across several representative vulnerability patterns, with recall exceeding 0.78 on SC2, E1, E2, and P4. Its performance varies more on some patterns, with P1, E3, E4, and SC3 showing lower recall and representing more challenging cases for \tool. Skill-Vetter and AIG attain high recall across many patterns, but their greater sensitivity results in substantially higher false-positive rates on the complete benchmark. In contrast, other scanners exhibit substantially lower recall across multiple patterns, resulting in more missed malicious Skills.

\subsection{Performance across Package Characteristics}
\label{subsec:package_characteristics}

We next examine how Skill package characteristics affect malicious-skill detection. We consider two dimensions: package size, measured by the number of files, and programming-language characteristics, including both language complexity and specific programming languages.

\begin{figure*}[t]
    \centering

    % Shared legend
    \includegraphics[width=\textwidth]{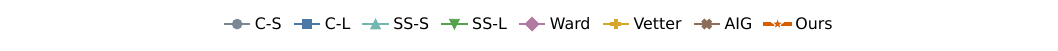}
    % \vspace{2mm}

    \begin{subfigure}[t]{0.32\textwidth}
        \centering
        \includegraphics[width=\linewidth]{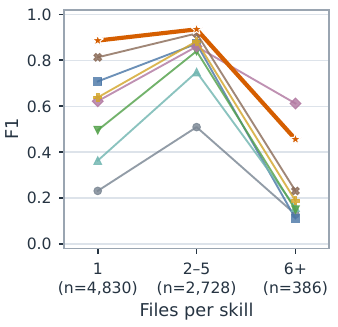}
        \caption{MalSkillBench}
        \label{fig:package_size_malskills}
    \end{subfigure}
    \hfill
    \begin{subfigure}[t]{0.32\textwidth}
        \centering
        \includegraphics[width=\linewidth]{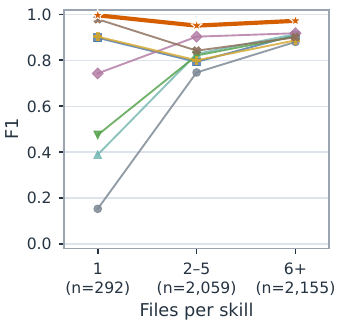}
        \caption{SkillTrustBench}
        \label{fig:package_size_skilltrust}
    \end{subfigure}
    \hfill
    \begin{subfigure}[t]{0.32\textwidth}
        \centering
        \includegraphics[width=\linewidth]{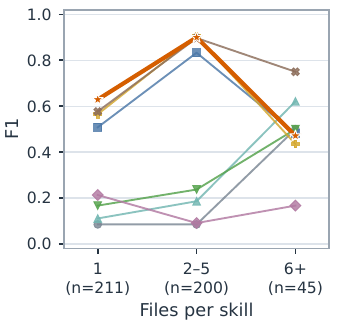}
        \caption{MaliciousAgentSkillsBench}
        \label{fig:package_size_masb}
    \end{subfigure}

    \caption{
        Detection performance across Skill package sizes.
    }
    \label{fig:package_size}
\end{figure*}

\noindent\textbf{Package size.}
We group Skills by the number of files in each package after excluding dataset-specific metadata and evaluation artifacts. As shown in Figure~\ref{fig:package_size}, \tool maintains strong F1 across the major package-size groups, achieving better overall performance than the baselines. On SkillTrustBench, \tool consistently outperforms all baselines across the three package-size groups. On MalSkillBench, it also achieves the strongest performance on the two dominant groups. Baselines exhibit substantially larger performance variations across package sizes. The pattern is less consistent on MASB. \tool achieves strong performance on single-file packages and remains competitive on packages containing two to five files, but its performance decreases on larger packages. Baselines show similar variations across package sizes, indicating that performance on MASB is more sensitive to package characteristics.

\begin{table*}[t]
\centering
\caption{
F1-score across packages with different programming-language complexity.
}
\label{tab:language_complexity}
\resizebox{\textwidth}{!}{
\begin{tabular}{llcccccccc}
\toprule
\textbf{Dataset} & \textbf{Group}
& \textbf{C-S} & \textbf{C-L}
& \textbf{SS-S} & \textbf{SS-L}
& \textbf{Ward} & \textbf{Vetter}
& \textbf{AIG} & \textbf{\tool} \\
\midrule

\multirow{3}{*}{MalSkillBench}
& No detected code   & .171 & .657 & .310 & .467 & .679 & .591 & .802 & \textbf{.819} \\
& One language       & .246 & .717 & .428 & .544 & .670 & .664 & .823 & \textbf{.911} \\
& Multiple languages & .484 & .830 & .683 & .771 & .816 & .827 & .880 & \textbf{.919} \\

\midrule

\multirow{3}{*}{SkillTrustBench}
& No detected code   & .364 & .372 & .571 & .750 & .545 & .279 & .571 & \textbf{.800} \\
& One language       & .692 & .773 & .736 & .716 & .848 & .769 & .827 & \textbf{.952} \\
& Multiple languages & .819 & .892 & .884 & .883 & .912 & .882 & .901 & \textbf{.970} \\

\midrule

\multirow{3}{*}{MASB}
& No detected code   & .000 & .263 & .000 & .100 & .000 & .368 & .400 & \textbf{.455} \\
& One language       & .034 & .815 & .109 & .205 & .087 & .863 & .880 & \textbf{.925} \\
& Multiple languages & .333 & .591 & .474 & .395 & .279 & .634 & \textbf{.667} & .471 \\

\bottomrule
\end{tabular}
}
\end{table*}

\begin{table*}[t]
\centering
\caption{
F1-score across programming languages. Language groups are non-exclusive, as a Skill may contain multiple programming languages.
}
\label{tab:language_performance}
\resizebox{\textwidth}{!}{
\begin{tabular}{llcccccccc}
\toprule
\textbf{Dataset} & \textbf{Language}
& \textbf{C-S} & \textbf{C-L}
& \textbf{SS-S} & \textbf{SS-L}
& \textbf{Ward} & \textbf{Vetter}
& \textbf{AIG} & \textbf{\tool} \\
\midrule

\multirow{9}{*}{MalSkillBench}
& Go         & .483 & .905 & .629 & .744 & .750 & .756 & .850 & \textbf{.919} \\
& JavaScript & .331 & .551 & .448 & .514 & .769 & .528 & .665 & \textbf{.845} \\
& PowerShell & .754 & .755 & .810 & .857 & .789 & .750 & .839 & \textbf{.874} \\
& Python     & .476 & .883 & .723 & .810 & .843 & .880 & .916 & \textbf{.934} \\
& Rust       & .378 & .852 & .609 & .821 & .857 & .828 & .877 & \textbf{.885} \\
& SQL        & .611 & .911 & .570 & .689 & .688 & .900 & .902 & \textbf{.939} \\
& Shell      & .402 & .760 & .583 & .673 & .751 & .737 & .843 & \textbf{.916} \\
& TypeScript & .464 & .818 & .610 & .730 & .764 & .766 & .864 & \textbf{.891} \\
& YAML       & .361 & .821 & .597 & .717 & .747 & .832 & .883 & \textbf{.891} \\

\midrule

\multirow{7}{*}{SkillTrustBench}
& JavaScript & .784 & .824 & .902 & .885 & .886 & .848 & .880 & \textbf{.959} \\
& PowerShell & .852 & .865 & .890 & .857 & .893 & .819 & .864 & \textbf{.930} \\
& Python     & .840 & .908 & .906 & .904 & .922 & .892 & .906 & \textbf{.974} \\
& SQL        & .687 & .725 & .878 & .818 & .866 & .819 & .805 & \textbf{.901} \\
& Shell      & .791 & .861 & .843 & .840 & .894 & .851 & .884 & \textbf{.965} \\
& TypeScript & .812 & .890 & .877 & .871 & .919 & .797 & .873 & \textbf{.932} \\
& YAML       & .735 & .846 & .804 & .805 & .899 & .861 & .884 & \textbf{.974} \\

\midrule

\multirow{5}{*}{MASB}
& JavaScript & .267 & \textbf{.766} & .474 & .450 & .222 & .818 & .844 & .452 \\
& Python     & .286 & .444 & .435 & .357 & .200 & .421 & .564 & \textbf{.593} \\
& Shell      & .125 & .763 & .241 & .282 & .148 & .822 & .839 & \textbf{.851} \\
& TypeScript & .400 & .444 & .333 & .286 & .500 & \textbf{.667} & .571 & .500 \\
& YAML       & .000 & .353 & .182 & .133 & .000 & \textbf{.588} & .429 & .444 \\

\bottomrule
\end{tabular}
}
\end{table*}

\noindent\textbf{Programming-language characteristics.}
We first examine performance under different levels of programming-language complexity. As shown in Table~\ref{tab:language_complexity}, \tool achieves strong F1 on both single- and multi-language packages in MalSkillBench and SkillTrustBench, outperforming all baselines in these groups. Its performance remains stable when moving from single-language to multi-language packages, indicating robustness to language heterogeneity. Packages without detected code are more challenging, suggesting that less code-oriented Skills provide fewer explicit implementation signals for security analysis.

We further examine whether detection performance varies across specific programming languages. As shown in Table~\ref{tab:language_performance}, \tool achieves the highest F1 across most language groups in MalSkillBench and SkillTrustBench, covering diverse implementations such as Python, Shell, JavaScript, and YAML. On MASB, \tool shows greater performance variation across programming languages, achieving the highest F1 on Python and Shell. Overall, \tool demonstrates strong detection performance across diverse programming languages.

\section{Complete Cross-Model Results}
\label{sec:backbone_results}

We provide the complete cross-model results underlying the analysis in
Section~\ref{subsec:llm_comparison} in Tables~\ref{tab:backbone_malskill}, \ref{tab:backbone_skilltrust}, and \ref{tab:backbone_masb}. To ensure reliable and fair efficiency comparisons, latency is measured under a controlled evaluation setting with consistent runtime configurations across models and methods.  Across the three benchmarks, direct zero-shot auditing exhibits high precision but lower recall, indicating that compact LLMs struggle to reliably identify malicious Skills. \tool improves malicious Skill detection across all evaluated backbones. It can recover malicious Skills missed under direct zero-shot auditing. The magnitude of improvement varies across backbones, indicating that the effectiveness of \tool remains influenced by the capability of the underlying compact LLM. Although these improvements introduce additional inference latency, the overall runtime remains practical, reflecting an effectiveness--efficiency trade-off.

\begin{table*}[t]
\centering
\caption{Complete cross-model results on MalSkillBench.}
\label{tab:backbone_malskill}
\small
\setlength{\tabcolsep}{4pt}
\begin{tabular}{llccccccc}
\toprule
\textbf{Model} & \textbf{Method}
& \textbf{Acc}
& \textbf{Prec}
& \textbf{Recall}
& \textbf{F1}
& \textbf{FPR}$\downarrow$
& \textbf{FNR}$\downarrow$
& \textbf{Latency (s)}$\downarrow$ \\
\midrule

\multirow{2}{*}{Gemma4 E4B}
& Zero-shot
& 0.737 & 0.992 & 0.474 & 0.641 & 0.004 & 0.526
& 10.2 \\
& \tool
& 0.909 & 0.944 & 0.869 & \textbf{0.905} & 0.051 & 0.132
& 27.4 \\

\multirow{2}{*}{Qwen3.5 9B}
& Zero-shot
& 0.811 & 0.990 & 0.626 & 0.767 & 0.007 & 0.374
& 31.1 \\
& \tool
& 0.918 & 0.969 & 0.862 & \textbf{0.912} & 0.027 & 0.138
& 62.1 \\

\multirow{2}{*}{DeepSeek-R1 8B (Thinking)}
& Zero-shot
& 0.734 & 0.972 & 0.479 & 0.641 & 0.014 & 0.521
& 11.7 \\
& \tool
& 0.806 & 0.925 & 0.664 & \textbf{0.773} & 0.054 & 0.336
& 21.8 \\

\multirow{2}{*}{DeepSeek-R1 8B (w/o Thinking)}
& Zero-shot
& 0.702 & 0.928 & 0.434 & 0.592 & 0.033 & 0.566
& 2.5 \\
& \tool
& 0.801 & 0.782 & 0.829 & \textbf{0.805} & 0.229 & 0.171
& 11.5 \\

\multirow{2}{*}{Mixtral}
& Zero-shot
& 0.531 & 0.983 & 0.057 & 0.108 & 0.001 & 0.943
& 8.3 \\
& \tool
& 0.629 & 0.838 & 0.312 & \textbf{0.455} & 0.060 & 0.688
& 14.2 \\

\bottomrule
\end{tabular}
\end{table*}

\begin{table*}[t]
\centering
\caption{Complete cross-model results on SkillTrustBench.}
\label{tab:backbone_skilltrust}
\small
\setlength{\tabcolsep}{4pt}
\begin{tabular}{llccccccc}
\toprule
\textbf{Model} & \textbf{Method}
& \textbf{Acc}
& \textbf{Prec}
& \textbf{Recall}
& \textbf{F1}
& \textbf{FPR}$\downarrow$
& \textbf{FNR}$\downarrow$
& \textbf{Latency (s)}$\downarrow$ \\
\midrule

\multirow{2}{*}{Gemma4 E4B}
& Zero-shot
& 0.712 & 0.873 & 0.640 & 0.739 & 0.163 & 0.360
& 11.8 \\
& \tool
& 0.957 & 0.971 & 0.961 & \textbf{0.966} & 0.050 & 0.039
& 28.5 \\

\multirow{2}{*}{Qwen3.5 9B}
& Zero-shot
& 0.858 & 0.996 & 0.779 & 0.875 & 0.005 & 0.221
& 36.3 \\
& \tool
& 0.959 & 0.982 & 0.952 & \textbf{0.967} & 0.030 & 0.048
& 65.7 \\

\multirow{2}{*}{DeepSeek-R1 8B (Thinking)}
& Zero-shot
& 0.637 & 0.990 & 0.432 & 0.602 & 0.007 & 0.568
& 15.2 \\
& \tool
& 0.792 & 0.945 & 0.715 & \textbf{0.814} & 0.072 & 0.285
& 22.9 \\

\multirow{2}{*}{DeepSeek-R1 8B (w/o Thinking)}
& Zero-shot
& 0.482 & 0.907 & 0.205 & 0.335 & 0.037 & 0.795
& 3.6 \\
& \tool
& 0.817 & 0.816 & 0.918 & \textbf{0.864} & 0.360 & 0.082
& 12.3 \\

\multirow{2}{*}{Mixtral}
& Zero-shot
& 0.373 & 0.929 & 0.014 & 0.027 & 0.002 & 0.986
& 6.4 \\
& \tool
& 0.535 & 0.904 & 0.301 & \textbf{0.451} & 0.056 & 0.699
& 17.1 \\

\bottomrule
\end{tabular}
\end{table*}

\begin{table*}[t]
\centering
\caption{Complete cross-model results on MASB.}
\label{tab:backbone_masb}
\small
\setlength{\tabcolsep}{4pt}
\begin{tabular}{llccccccc}
\toprule
\textbf{Model} & \textbf{Method}
& \textbf{Acc}
& \textbf{Prec}
& \textbf{Recall}
& \textbf{F1}
& \textbf{FPR}$\downarrow$
& \textbf{FNR}$\downarrow$
& \textbf{Latency (s)}$\downarrow$ \\
\midrule

\multirow{2}{*}{Gemma4 E4B}
& Zero-shot
& 0.656 & 0.500 & 0.038 & 0.071 & 0.020 & 0.962
& 9.2 \\
& \tool
& 0.878 & 0.892 & 0.737 & \textbf{0.807} & 0.048 & 0.263
& 19.4 \\

\multirow{2}{*}{Qwen3.5 9B}
& Zero-shot
& 0.669 & 1.000 & 0.038 & 0.074 & 0.000 & 0.962
& 27.1 \\
& \tool
& 0.831 & 0.857 & 0.612 & \textbf{0.714} & 0.054 & 0.389
& 65.5 \\

\multirow{2}{*}{DeepSeek-R1 8B (Thinking)}
& Zero-shot
& 0.675 & 1.000 & 0.057 & 0.108 & 0.000 & 0.943
& 10.5 \\
& \tool
& 0.827 & 0.820 & 0.637 & \textbf{0.717} & 0.074 & 0.363
& 22.7 \\

\multirow{2}{*}{DeepSeek-R1 8B (w/o Thinking)}
& Zero-shot
& 0.675 & 0.714 & 0.096 & 0.169 & 0.020 & 0.905
& 2.2 \\
& \tool
& 0.779 & 0.639 & 0.822 & \textbf{0.719} & 0.244 & 0.178
& 15.0 \\

\multirow{2}{*}{Mixtral}
& Zero-shot
& 0.656 & 0.000 & 0.000 & 0.000 & 0.000 & 1.000
& 3.9 \\
& \tool
& 0.772 & 0.730 & 0.535 & \textbf{0.618} & 0.104 & 0.465
& 11.8 \\

\bottomrule
\end{tabular}
\end{table*}

\section{Prompt Templates}
\label{sec:prompt_templates}

We provide the prompt templates used by the compact LLM in \tool. The LLM performs two semantic reasoning tasks: intent analysis and evidence-guided risk adjudication. The templates below present the core instructions and output structures used in our implementation.

\subsection{Intent Analysis Prompt}

\begin{promptbox}{Intent Analysis}

\small
\ttfamily

[INPUT]

Skill Name: \{skill\_name\}

Skill Specification: \{skill\_md\}

[INSTRUCTION]

Infer the intended functionality of the Skill.

1. Identify its declared purpose.

2. Identify the capabilities necessary and
   proportionate to that purpose.

[OUTPUT]

\{
  "declared\_purpose": "...",

  "expected\_capabilities": [...]
\}

\end{promptbox}

\begin{promptbox}{Evidence-Guided Risk Adjudication}

\small
\ttfamily

[INPUT]

Skill Specification: \{skill\_md\}

Functional Specification: \{intent\_analysis\}

Evidence Report: \{evidence\_report\}

[INSTRUCTION]

Assess the Skill using the grounded evidence
and its intended functionality.

1. Review suspicious evidence and its source.

2. Assess whether each behavior is necessary
   for the declared functionality.

3. Identify direct benign counterevidence.

4. Determine whether the evidence forms an
   abuse chain and make the final judgment.

[DECISION POLICY]

- Security-sensitive capability alone does not
  imply maliciousness.

- Necessary and proportionate behavior should
  normally be treated as benign.

- Strong unexplained or deceptive behavior
  supports a malicious judgment.

- Critical agentic risks require direct,
  risk-specific benign counterevidence.

[OUTPUT]

\{
  "suspicious\_evidence\_review": [...],

  "benign\_counterevidence": [...],

  "abuse\_chain": \{
    "present": true/false,
    "components": [...]
  \},

  "is\_malicious": true/false,

  "confidence\_score": 0--100,

  "reasoning": "..."
\}

\end{promptbox}

\end{document}